\documentclass[prl,aps,onecolumn,floatfix]{revtex4}
\usepackage{graphicx,amssymb,amsmath,color,psfrag}
\usepackage{amsthm}
\usepackage{amsfonts}
\usepackage{algorithmic}
\usepackage{enumerate}
\usepackage{latexsym}

\usepackage{color}

\usepackage{ulem}

\begin{document}
\newcommand{\dt}{\Delta\tau}
\newcommand{\al}{\alpha}
\newcommand{\ep}{\varepsilon}
\newcommand{\ave}[1]{\langle #1\rangle}
\newcommand{\have}[1]{\langle #1\rangle_{\{s\}}}
\newcommand{\bave}[1]{\big\langle #1\big\rangle}
\newcommand{\Bave}[1]{\Big\langle #1\Big\rangle}
\newcommand{\dave}[1]{\langle\langle #1\rangle\rangle}
\newcommand{\bigdave}[1]{\big\langle\big\langle #1\big\rangle\big\rangle}
\newcommand{\Bigdave}[1]{\Big\langle\Big\langle #1\Big\rangle\Big\rangle}
\newcommand{\braket}[2]{\langle #1|#2\rangle}
\newcommand{\up}{\uparrow}
\newcommand{\dn}{\downarrow}
\newcommand{\bb}{\mathsf{B}}
\newcommand{\ctr}{{\text{\Large${\mathcal T}r$}}}
\newcommand{\sctr}{{\mathcal{T}}\!r \,}
\newcommand{\btr}{\underset{\{s\}}{\text{\Large\rm Tr}}}
\newcommand{\lvec}[1]{\mathbf{#1}}
\newcommand{\gt}{\tilde{g}}
\newcommand{\ggt}{\tilde{G}}
\newcommand{\jpsj}{J.\ Phys.\ Soc.\ Japan\ }

\title{Controllable magnetic correlation between two impurities by spin-orbit coupling in graphene}
\author{F. M. Hu*}
\affiliation{Bremen Center for Computational Materials Science, University of Bremen, Am Fallturm 1a, D-28359 Bremen, Germany}
\author{Liangzhi Kou*}
\affiliation{Bremen Center for Computational Materials Science, University of Bremen, Am Fallturm 1a, D-28359 Bremen, Germany}
\author{Thomas Frauenheim}
\affiliation{Bremen Center for Computational Materials Science, University of Bremen, Am Fallturm 1a, D-28359 Bremen, Germany}

\begin{abstract}
Two magnetic impurities on the edge of a zigzag graphene nanoribbon strongly interact with each other via indirect coupling, which can be mediated by conducting carriers. By means of Quantum Monte Carlo (QMC) simulations, we find that the spin-orbit coupling $\lambda$ and the chemical potential $\mu$ in system can be used to drive the transition of local-spin exchange from ferromagnetism to anti-ferromagnetism. Since the tunable ranges for $\lambda$ and $\mu$ in graphene are experimentally reachable, we thus open the possibilities for its device application. The symmetry in spatial distribution is broken by the vertical and the transversal spin-spin correlations due to the effect of spin-orbit coupling, leading to the spatial anisotropy of spin exchange, which distinguish our findings from the case in normal Fermi liquid.

Correspondence and requests for materials should be addressed to F. M. Hu (feiminghu@gmail.com) or Liangzhi Kou (kouliangzhi@gmail.com)
\end{abstract}

\maketitle

Due to the unusual electronic quasi-particles and the flexible carrier concentration \cite{RMP09}, graphene with local spins is regarded as a novel type of two-dimensional magnetic semiconductor as well as a promising candidate for spintronics devices \cite{NM12}. Doping with magnetic impurities is regarded as an effective route to induce local spins in graphene \cite{Fritz13}, because the pristine graphene is a nonmagnetic semimetal without band gap.  In previous works concerning single-impurity doped graphene, it was reported that applied gate voltage (to change the chemical potential) can control the electron filling on impurity orbit \cite{Bruno08,Hu12} and its Kondo screening from conducting electrons \cite{Wehling10,Bruno11}, thus leading to an on-off switching for a single spin in graphene. But in realistic spin-based devices, we need to consider multi-spin case and spin-flip process owing to the experimental challenge of doping concentration control,  which will induce spin current transport in graphene systems. Therefore, for the device application, it is crucial to achieve efficient controllability for spin correlations among embedded impurities, which is measured by the Ruderman-Kittel-Kasuya-Yosida (RKKY) interaction and mediated by conducting electrons. In previous studies  on RKKY interaction, the electron doping dominates the large-distance oscillation with a wave vector $2k_F$ as well as decay behavior $R^{-\gamma}$ \cite{Brey07}.

Recently, there is a surge of interest in effect of spin-orbit coupling (SOC) on electronic properties, which opens a class of novel time-reversal invariant topological insulators (TI) \cite{RMP11}. On the edge (two-dimension case) or surface (three-dimension case) of these TI materials, the carries have spin-momentum locking, so called "helical" states: electrons with opposite momenta have orthogonal spin orientations, so that the backscattering $k$ $\leftrightarrow$ $-k$ is forbidden without a spin flip. Due to this scenario, there is spin-current vortex around the single local spin on the helical edge\cite{Wu06}. Furthermore, on the surface of three-dimensional realistic TI, Kondo impurities will open up a local gap \cite{Liu09} and the exchanges can be described by the RKKY interaction among the impurities\cite{Zhu11}. More interestingly, the combination of anisotropic spin exchange and SOC on the edge of two-dimensional realistic TI results in unnormal temperature-dependent conductance on the edge \cite{Maciejko09,Tanaka11}. These investigations indicate that the SOC can greatly modify the spin configuration for single impurity in realistic TI, such as class of Bi$_{2}$Se$_{3}$ (Bi$_{2}$Te$_{3}$, Sb$_{2}$Te$_{3}$) \cite{Zhang09} and strained HgTe \cite{Konig07}. While in graphene, which is one of the earliest proposals for TI materials, multi-spin problem has not been comprehensively studied, despite the facts that the SOC in graphene can be enhanced in a wide range, while both the Kondo screening around single spin and RKKY interaction between two spins can be greatly affected by the SOC.

In this paper, we consider two Anderson impurities on the edge of a zigzag graphene ribbon, and investigated the effect of Kane-Mele SOC and chemical potential on the local spin exchange, which can be achieved by heavy metal doping or electric field. Through Quantum Monte Carlo simulation at finite temperature, we carry out a comprehensive study on the spin and charge coupling strengthes of this impurity pair. For the spin-spin correlation, which measures the RKKY interaction, we found that, the in-plane components of correlations favor ferromagnetism but the out-plane correlation can be tuned from ferromagnetism to antiferromagnetism by the SOC, while such transition is much dependent on the on-site Coulomb interaction of impurity orbital. From recent first-principal and experimental results of enhanced SOC in graphene, we conclude that spin-spin exchange among impurities in graphene is experimentally tunable by SOC. In contrast, the charge fluctuation between two impurities is always suppressed. Meanwhile, in-plane components of correlation will decay with an oscillation behavior when the distance between two impurities is increased, but the out-plane spin and charge correlations will decay smoothly, which is totally different from the Fermi liquid behavior, and leading to the spatial anisotropy of spin exchange.

\begin{figure}
\begin{center}
\includegraphics[scale=0.80, bb= 126 30 5 115]{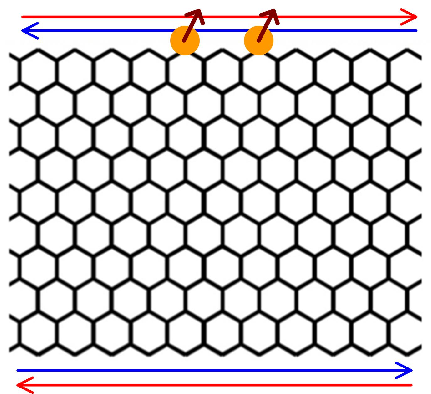}
\end{center}
\caption{(color online) The scheme for the two magnetic impurities (yellow balls) at the edge of the zigzag nanoribbon. The arrows on the top and bottom demotes the moving directions for two species spins.}\label{Fig:fig1}
\end{figure}

 To simulate the model, we construct a model with width of 20 {\AA} zigzag-edge graphene ribbon and along the horizontal direction we adopt periodical condition. As indicated in Kane-Mele investigation \cite{Kane05}, graphene will be driven into topological phase when the intrinsic SOC in graphene is enhanced. In the case, the ribbon would possess the helical edge states. In Fig.~\ref{Fig:fig1}, we show the scheme for two impurities (yellow balls) located on the zigzag helical edge, where red and blue arrows denote the conducting electrons with up and down spins, respectively, and they propagate along oppositive directions (spin-momentum locking).

The QMC simulation is based on two-Anderson-impurity model, the total Hamiltonian reads as
\begin{equation}\label{Eq:KMA}
H=H_{\textrm{K-M}}+H_{1}+H_{2},
\end{equation}
where
$H_{\textrm{K-M}}$ is the Kane-Mele model in a zigzag edge graphene ribbon \cite{Kane05}.  It is composed with two parts:
$H_{\textrm{K-M}}=H_{t}+H_{\textrm{so}}$,
with $H_{t}$ being the usual nearest-neighbor hopping of tight-binding model in graphene
\[
H_{t}=-t\sum_{<ij>,\sigma}(c^{\dag}_{i,\sigma}c_{j,\sigma}+\textrm{H.c.})-\mu\sum_{i,\sigma}c^{\dag}_{i,\sigma}c_{i,\sigma},
\]
and
\[
H_{\textrm{so}}=\lambda\sum_{<<ij>>}(i\nu_{ij}c^{\dag}_i\sigma^zc_{j}+\textrm{H.c.})
\]
$H_{\textrm{so}}$ is the SOC term and $\sigma^{z}$ is the $z$ Pauli matrix. $H_{\textrm{so}}$ thus has opposite signs for opposite electron spins.
The parameters $\nu_{ij}=-\nu_{ji}=\pm1$ depend on the orientation of the two nearest neighbor bonds as the electron hops from site $i$ to $j$: $\nu_{ij}=+1$ if the
electron makes a left turn to the second bond. It is negative if it makes a right turn.
$H_1$ is the part for two impurities
\[
H_{1}=\sum_{\sigma m=1,2}(\varepsilon_{m}-\mu)d^{\dag}_{\sigma m}d_{\sigma m}^{}+U_m d^{\dag}_{\uparrow m}d_{\uparrow m}^{}d^{\dag}_{\downarrow m}d_{\downarrow m}^{}\texttt{.}
\]
and $H_2$ describes the hybridization between impurities and conducting electrons
\[
H_{2}=\sum_{\sigma m=1,2}V_{m}[c^{\dag}_{m\sigma}d_{\sigma m}^{}+d^{\dag}_{\sigma m}c_{m\sigma}^{}]\texttt{.}
\]
Here $m$ is the index of two impurities, in $H_1$ and $H_2$, $\varepsilon_{m}$ is the energy of the impurity orbital and $U_m$ is Coulomb repulsion  inhibiting the simultaneous occupancy of the orbital by two electrons and $V_m$ is the hybridization strength. In this work, we consider the two  identical impurities, so the above parameters in Hamiltonian are $m$-independent.

We use the single-impurity QMC base on discrete-time algorithm \cite{Hirsch86} to compute the local thermodynamic properties of the impurity. The length of imaginary-time is set as $\Delta\tau = 0.1$ to suppress the systemic error. The QMC naturally returns the imaginary-time Green's function $G_{d\sigma}(\tau > 0)= \langle d_{\sigma}(\tau)d_{\sigma}^{\dag}\rangle $ of the impurity. With this Green's function, we can compute the spin-spin correlations for two impurities along $\alpha=x,y,z$ directions versus distance as
\[
S^{\alpha}(R)=\langle s_1^{\alpha}s_2^{\alpha}\rangle,
\]
where $s_m^{x}=(d^{\dag}_{\uparrow m}d_{\downarrow m}+d^{\dag}_{\downarrow m}d_{\uparrow m})$,
$s_m^{y}=(-i)^2(d^{\dag}_{\uparrow m}d_{\downarrow m}-d^{\dag}_{\downarrow m}d_{\uparrow m})$ and
$s_m^{z}=(d^{\dag}_{\uparrow m}d^{}_{\uparrow m}-d^{\dag}_{\downarrow m}d^{}_{\downarrow m})$;
as well as the charge-charge correlation as the function of distance
\[
C(R)=\langle n_1n_2 \rangle-\langle n_1\rangle \langle n_2 \rangle,
\]
and $n_m=(d^{\dag}_{\uparrow m}d^{}_{\uparrow m}+d^{\dag}_{\downarrow m}d^{}_{\downarrow m})$.
The spin-spin correlations $S^{\alpha}(R)$ directly reflect the RKKY interaction which is mediated by conducting electrons on the helical edge, while charge-charge correlation $C(R)$ indicates the indirect Coulomb repulsion.

\begin{figure}
\begin{center}
\includegraphics[scale=0.50, bb=18 16 526 436]{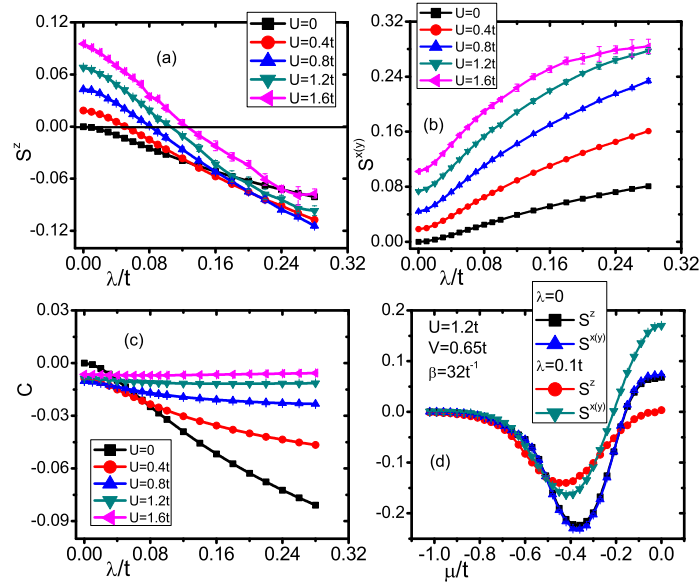}
\end{center}
\caption{(color online) (a) The spin-spin correlation along z direction $S^{z}$ versus $\lambda$, (b) The spin-spin correlation along x or y direction $S^{x(y)}$ versus $\lambda$ (c) The charge-charge correlation versus $\lambda$ and in (a)-(c) the inverse of temperature $\beta=32t^{-1}$,
$V=0.65t$, $\mu=0$ and $\varepsilon=-\frac{U}{2}$. (d) $S^{z}$ and $S^{x(y)}$ versus chemical potential $\mu$ with $\beta=32t^{-1}, U=1.2t, V=0.65t$ and $\varepsilon=-\frac{U}{2}$.}\label{Fig:fig_mu_soc}
\end{figure}

In Fig.~\ref{Fig:fig_mu_soc}, we show the spin-spin and charge-charge correlations as function of SOC, Coulomb interaction, and chemical potential. Here, we fixed the hole-particle symmetry $\mu=0$ and $\varepsilon=-\frac{U}{2}$, so that $\langle d^{\dag}_{\uparrow m}d^{}_{\uparrow m}+d^{\dag}_{\downarrow m}d^{}_{\downarrow m}\rangle=1$ and the local moments at two impurity sites are well developed. One can see from Fig.~\ref{Fig:fig_mu_soc}(a), the out-plane correlations of $S^{z}$ highly depend on SOC, which can be reduced in a linearly way and have a wide variation range.  When SOC is increased, the value of vertical correlation $S^{z}$ can be changed from positive to negative, it means that a transition from a ferromagnetic (FM) to anti-ferromagnetic (AFM) correlation is achieved when the SOC reaches a critical value. It is interesting to note that, the critical SOC strength for such a transition increases as Coulomb interaction is increased, from the QMC simulations. For example, the critical SOC value is about 100 meV when $U$ is 1eV, but it is increased to 300 meV when $U$ is is 5eV. The underlying mechanism can be attributed to the competition of spin and charge fluctuations between two impurities. When they are close to each other, the changing in carriers SOC can directly result in the spin-flip process which reflects in $S^{z}$, but the increasing Coulomb interaction $U$ can block spin flip, so it is seen that with a larger $U$ in Fig.~\ref{Fig:fig_mu_soc}(a), the conducting electrons need larger kinetic energies (SOC) to drive a FM-AFM transition. But when two impurities are far away from with each other, we can not see this transition in $S^{z}$ but in $S^{x(y)}$, which is totally determined by the Luttinger liquid physics on the edge \cite{Hohenadler12}. The explanation can be also demonstrated in the latter plots [Fig. 3(a)].

The SOC driven magnetism transition will be important for the spintronics application. It is worthy to point out the required critical SOC is in a reasonable range based on the fact that recent extensive strategies were proposed and developed to enhance SOC in graphene. In Table~\ref{Tab:soc}, we sumarize the recent achievements from both density functional theory and experiments. In the case of interface/intercalation, the recent tunable range for SOC is 50 - 225 meV through placing graphene on the substrates of Au(111), Ag(111), and Ni(111) and intercalation with SiC (0001)\cite{Nicolay01,Dedkov08,Li11,Marchenko12,Li13}. Alternatively, doping graphene with heavy or magnetic adatoms (Ni, Au, Ti and Os) is also able to induced the enhanced SOC in graphene of 20 - 500 meV \cite{Abdelouahed10,Weeks11,HuJun12}. Meanwhile, when sandwiching graphene between strong spin orbit interactions or Rashba materials by the proximity effects, the intrinsic SOC is significantly enhanced and graphene is driven into topological phase with nontrivial gap of 120 meV \cite{Kou2013, arxiv2014}. Due to these efficient strategies, it is possible to tune the SOC in graphene over a wide range, the FM-AFM transition for spin correlation between two magnetic impurities via SOC controlling is thus feasible.


In contrast to the spin-spin correction transition of $S^{z}$, the transversal correlation $S^{x(y)}$ is increased linearly by the SOC interaction but not change its sign. In Fig.~\ref{Fig:fig_mu_soc}(b), we see that at SOC $\lambda=0$, $S^{z}=S^{x(y)}$, but this symmetry is broken when SOC is switched on. The transversal correlation $S^{x(y)}$ is increased when either of SOC and Coulomb interaction is enhanced. More interestingly, the enhance SOC not only can change the spin correction between two impurities, but also affect the charge interaction between them. As indicated in Fig.~\ref{Fig:fig_mu_soc}(c), the charge-charge correlation as a function of SOC with different values of Coulomb interaction is calculated. We can see that the absolute value of $C$ is increased when SOC is increased, but decreased as the Coulomb interaction is enhanced. When $U$ is at $1.2t$, the charge fluctuation is totally suppressed. To have a comprehensive physics picture, we also vary chemical potential $\mu$ to break the hole-particle symmetry and study their effects on $S^{z}$ and $S^{x(y)}$, In Fig.~\ref{Fig:fig_mu_soc}(d). The vertical and transversal correlations are totally degenerated ($i.e.$ $S^{z}$ is equal to $S^{x(y)}$) at SOC strength $\lambda=0$. But when $\lambda\neq0$, the SOC effect on both spin correlation is different, leading to non-degenerated $S^{z}$ and $S^{x(y)}$. We then turn to the effect of chemical potential on $S^{z}$ and $S^{x(y)}$, it could be found that both $S^{z}$ and $S^{x(y)}$ are positive near the zero point of $\mu$, suggesting that two impurities correlate ferromagnetically. After $\mu$ is shifted from 0, the two impurities have anti-ferromagnetic correlation as indicated by the negative values of $S^{z}$ and $S^{x(y)}$ ( $\mu$ is in the range of -0.2$t$ to -0.8$t$). But when $\mu$ is further reduced, both $S^{z}$ and $S^{x(y)}$ approach to zero, indicating the weak coupling between two impurities.



\begin{figure}
\begin{center}
\includegraphics[scale=0.5, bb=30 59 476 399]{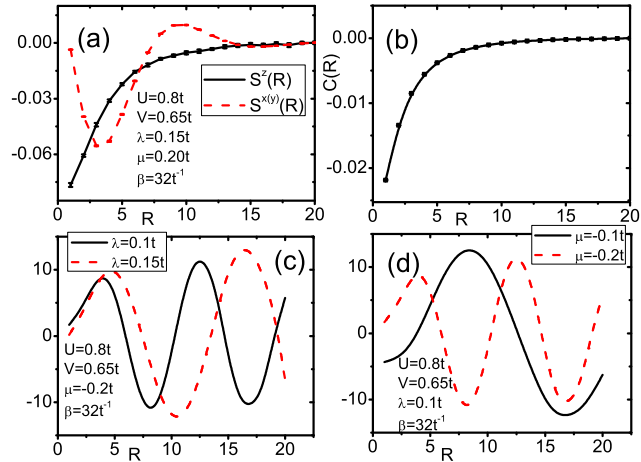}
\end{center}
\caption{(color online) (a) The vertical and transversal spin-spin correlation as function of distance between two impurities (b) The charge-charge correlation as function of distance. (c) The ratio $S^{x(y)}(R)/C(R)$ with fixed chemical potential $\mu=-0.2t$ and varied SOC $\lambda=0.1t$ and $0.15t$. (d) The ratio $S^{x(y)}(R)/C(R)$ with fixed SOC $\lambda=0.1t$ and varied chemical potential $\mu=0.1t$ and $0.2t$.The $x$ axis is in the unit of lattice constant of graphene.}\label{Fig:fig_sz_sxy_asy}
\end{figure}

The distance between two impurities is expected to influence the couplings of spin-spin and charge-charge due to the fact that the RKKY interation is sensitive to the distance. We preformed the QMC calculations to address the problem when varying the distance. In Fig.~\ref{Fig:fig_sz_sxy_asy}, we show the spin-spin and charge-charge correlations as functions of distance between two impurities. Here we consider the case that both of two impurities are located on the edge (the same sub-lattice). In Fig.~\ref{Fig:fig_sz_sxy_asy}(a), we see that the vertical spin-spin correlation $S^{z}(R)$ and transversal one $S^{x(y)}(R)$ have different spatial distribution. $S^{x(y)}(R)$ has both the behavior of decaying and oscillating, but $S^{z}(R)$ is always decaying with the increased distance. Similar to $S^{z}(R)$, the charge-charge correlation is also decaying with distance (Fig.~\ref{Fig:fig_sz_sxy_asy}(b)). These correlation functions have the following behavior:
\[
S^{z}(R) \propto \frac{1}{f_1(R)},
\]
\[
S^{x(y)}(R) \propto \frac{\cos(\phi R)}{f_2(R)}
\]
and
\[
C(R) \propto \frac{1}{f_3(R)},
\]
here $f_1(R)$, $f_2(R)$ and $f_3(R)$ are decaying function of distance has the formula as $1/R^{\gamma}$.

To more clearly display these correlations, we calculate the ratios $S^{x(y)}(R)/C(R)$ and $S^{z}(R)/C(R)$. We find that $S^{z}(R)/C(R)$ is a trivial constant, but $S^{x(y)}(R)/C(R)$ is an oscillating function of distance, as shown in Fig.~\ref{Fig:fig_sz_sxy_asy} (c) and (d). The ratio exhibits prefect behavior of sine or cosine functions with wave vector, which depends on the chemical potential and SOC.


Finally, we discuss the model Hamiltonian of Eq.~(\ref{Eq:KMA}) in more realistic situations with the effect of Hubbard $U'$ in graphene and hybridization. In realistic graphene zigzag nanoribbon, the Coulomb interaction plays a signification role and stabilizes a spin-polarized ground state \cite{Son06}. In this case, the conducting electron at the edge can be regarded as standard one-dimensional Luttinger liquid, coupled with which the two magnetic impurities at low temperature have the RKKY interaction as the asymptotic behavior $R^{-\gamma}\cos2k_FR$, here $k_F$ is the Fermi momentum. The decaying factor $\gamma$ is the function of Hubbard $U'$: when $U'=0$, $\gamma=1$, and
when $U'\neq0$, $0<\gamma<1$ \cite{Egger96,Hallberg97}. So in Luttinger liquid case, at low temperatures the RKKY interaction decays more slowly  than the case in Fermi liquid $U'=0$ . The methods used to enhance the graphene's SOC in Table~\ref{Tab:soc} are based in the mechanism that via the hybridization between the electronic bands of carbon atoms and heavy atoms with large intrinsic SOC, the SOC of graphene will be greatly enhanced. The hybridization depends on three parameters of heavy atoms: the crystal field, the intrinsic SOC and the hopping between carbon and external atoms. If the crystal field is greatly larger than the later two quantities, the graphene $\pi$ bands still dominate near the Fermi energy and Kane-Mele Hamiltonian is also valid and the effect of hybridization can be absorbed in the enhanced SOC through second order approximation \cite{Weeks11} and furthermore when the crystal field decreases and the intrinsic SOC and the hopping between carbon and external atoms increase, the SOC in graphene will increase. While if these three parameters have the same order and crystal field is close to Dirac point, on one hand such strong hybridization can further enhance the SOC in graphene and influence the RKKY interaction as shown in Fig.~\ref{Fig:fig_mu_soc}(a)-(c); on the other hand, there are additional bands near the Fermi energy from heavy atoms, the electrons from these bands will exchange with those at the levels of impurities (both spin and charge exchanging). As a result of this exchanging, the spin-spin and charge-charge correlations will increase. In addition, the three ways in Table~\ref{Tab:soc} can raise charge transferring and shift the Fermi energy of pristine graphene from Dirac point, so we do the calculation varying the chemical potential and show the effects in Fig.~\ref{Fig:fig_mu_soc}(d).

In conclusion, we investigate two Anderson impurities on a zigzag graphene nanoribbon by QMC simulation in present work. When the system is under a topological phase transition driven by Kane-Mele SOC, we find that , the spin rotation symmetry in RKKY interaction between two impurities can be partly broken because of the SOC. As the two impurities are close to each other and the Fermi energy is near the Dirac point, the in-plane RKKY interactions favor ferromagnetism but the out-plane component can be tuned from ferromagnetism to anti-ferromagnetism by the SOC. The distance effect on RKKY interaction is also systermatically investigated, the in-plane interaction has both behavior of decaying and oscillation, but the out-plane interaction and charge-charge correlation only has decaying as the distance between them is increased. This behavior is totally different from the conventional Fermi liquid. Our theoretical findings here provide some useful suggestions of the spin-polarized STM measurement to RKKY interaction\cite{Zhou10}.

\section{Author Contributions Statement}
F.H. conceived the idea and do the QMC calculation, L.K. made the first-principle discussion and collected the data in the Table \ref{Tab:soc}, and F.H. and L.K. wrote the manuscript together. F.H., L.K. and T.F. discussed the results and reviewed the manuscript.

\section{Additional information}
\textbf{Competing financial interests}: The authors declare no competing financial interests.

\newpage

\begin{table}[top]
\caption{Enhanced Gap $\Delta E_{\textrm{SOC}}$ in Graphene}
\label{schedule}
\centering %
\begin{tabular}{clc}
\toprule
Method                   &~~~~~                               $\Delta_{\textrm{SOC}}$ (meV) \\ \hline
Interface/intercalation  &~~~~~      50 - 225 \cite{Nicolay01,Dedkov08,Li11,Marchenko12,Li13}\\ \hline
Adatom doping            &~~~~~      20 - 500 \cite{Abdelouahed10,Weeks11,HuJun12}          \\ \hline
Proximity effect         &~~~~~      50 - 120 \cite{Kou2013, arxiv2014}                     \\ \hline\label{Tab:soc}
\end{tabular}
\end{table}

\end{document}